\documentstyle[prl,aps,graphicx,preprint]{revtex}

\begin{document}
\draft


\title{Charge Density Waves and $x=1/8$ anomaly in
La$_{2-x-y}$(Nd,Eu)$_y$(Ba,Sr)$_x$CuO$_4$}

\author{I.A. Larionov and M.V. Eremin}

\address{Magnetic Radiospectroscopy Laboratory, Kazan State University,
420008 Kazan, Russia}

\maketitle
\date{February 15, 2002}

\begin{abstract}
We adopt a $t_1-t_2-t_3-J-G$ model for explanation of $x=1/8$ anomaly in
La$_{2-x}$Sr$_x$CuO$_4$ family compound. The calculated charge
susceptibility shows a maximum near $(\pi,\pi)$ at intermediate temperatures
and near $(\pi,\pi/2)$ as temperature approaches zero, in agreement with
neutron scattering experiments. Coulomb repulsion $G$ between the first
neighbors turns out to be the source of Charge Density Waves (CDW) in narrow
band $t_1^{\mathit{eff}}, t_2^{\mathit{eff}}, t_3^{\mathit{eff}}<G$. For
physically realistic hopping values we obtain the CDW amplitude $e_Q=x$. The
in-phase domain structure as a candidate for "stripe" picture is proposed.

\end{abstract}

\pacs{74.72.Dn, 71.45.Lr, 74.72.-h, 71.27.+a}

It is widely accepted, that high-temperature superconductivity (HTSC)
appears in the vicinity of metal-insulator transition (MIT) in layered
copper oxides \cite{Imada}. In addition, these compounds possess a
pseudogap phenomenon in the normal state of underdoped regime, namely the
density of states suppression at Fermi level. The superconducting CuO$_2$
planes in La$_{2-x}$Sr$_x$CuO$_4$, YBa$_2$Cu$_3$O$_{7-y}$ and other copper
oxide layered HTSC compounds present nearly an ideal realization of
two-dimensional (2D) systems. Low dimensional character of these compounds
suggests, that various phases are possible due to instabilities. It is
known, that s-wave BCS Superconductivity compete with Charge Density Waves
(CDW), where the anomalous averages were due to Fr\"ohlich interaction and
did not depend on the wave vector. $d$-wave superconductivity was also
found to compete with $id$-wave Peierls instability in bilayered underdoped
HTSC, where both gaps were due to Superexhange and Coulomb interactions
between the nearest neighbors \cite{MyJETPL98}. Very recently this sliding
vortex-antivortex current structures in CuO$_2$ planes attracted a renewed
interest in connection with unusual nonsuperconducting state properties of
layered copper oxide HTSC compounds
\cite{MyJETPL98,EreminRigamonti,SudipChakra}.

Strong depression of T$_{\mathrm C}$ in La$_{2-x}$Ba$_x$CuO$_4$
\cite{Moodenbaugh}, La$_{2-x}$Sr$_x$CuO$_4$ \cite{Kumagai} and
La$_{1.48}$Nd$_{0.4}$Sr$_{0.12}$CuO$_4$ \cite{Crawford} within narrow region
of $x$ at $x\approx 1/8$ is the well known important problem of the doped
CuO$_2$ plane in the HTSC. The discovery of stripe like structure via
neutron diffraction in La$_{1.48}$Nd$_{0.4}$Sr$_{0.12}$CuO$_4$
\cite{Tranquada45} gave a new accelerating impulse for study this intriguing
problem (see \cite{TranquadaReview} for review). New ideas regarding the
relation between the fluctuating stripe motion and HTSC phenomenon were
speculated \cite{StripesSpec}. The ${\bf Q}$ dependence of peaks as seen in
neutron diffraction experiments seems to provide an evidence that stripes
are a kind of CDW instability; this scenario was considered in a number of
papers (see e.g. \cite{StripeTh1}). However, the crucial question, namely
why at so narrow interval of concentration around $x \approx 1/8$ the
mystery of $T_C$ strong depression occurs and accompanies by stripes pattern
phenomenon is not been completely understood. Recently this problem was
investigated by magnetic resonance method \cite{Teitelbaum} and new evidence
of the quasi-static charge-spins ordering at $x \approx 1/8$ has been found
at low temperatures in La$_{2-x}$Ba$_x$CuO$_4$ \cite{MatsumuraJpn} and in
La$_{1.48}$Eu$_{0.4}$Sr$_{0.12}$CuO$_4$ \cite{SawaJpn}. Note that x-ray
diffraction measurements have confirmed the existence of charge order in
La$_{1.48}$Nd$_{0.4}$Sr$_{0.12}$CuO$_4$ \cite{ZimmermannEur}.

In this Letter we will concentrate ourselves to the description of
properties for particular underdoped case $x = 1/8$ in
La$_{2-x}$Sr$_x$CuO$_4$ on the {\it microscopical} basis. We start from the
$t_1-t_2-t_3-J-G$ model Hamiltonian
\begin{eqnarray} \label{Hamiltonian}
 & \hat{H} & = \sum_{ij}t_{ij}^{}\Psi_i^{pd,\sigma}\Psi_j^{\sigma,dp}+
\sum_{l>m}J_{lm}\left[ \left( {\bf S}_l {\bf S}_m\right)-
\frac{n_l n_m}{4} \right]  \nonumber \\
 & + & \sum_{l>m}G_{lm} \delta_l\delta _m -\sum_{i} F_{i} {\big( }
 \sum_{\rho} S_{i+{\rho}} {\big )}^2 ,
\end{eqnarray}
where $\delta_l = 1- n_l $ is the number of {\it extra} holes per unit cell,
$\Psi_i^{pd,\sigma} (\Psi_j^{\sigma,dp})$ are creation (annihilation)
operators constructed on the singlet combination of copper (d) and oxygen
(p) states basis (see Ref. \onlinecite{EreminBand} for details). The last
term takes into account the polarization of nearest neighbor (n.n.) copper
spins ($\rho$ runs over n.n.) around copper-oxygen singlet at site $i$ as
described in \cite{EreminSigmund}.

Following the idea of Zhang and Rice\cite{ZhangRice} about copper-oxygen
singlets formation it was shown \cite{EreminBand}, that it is possible to
describe correctly the elementary excitations spectrum in cuprates. This
singlet correlated band is analogous to upper Hubbard band with essential
distinction - the subband splitting is much smaller, compared to Hubbard
model. Therefore it is possible to apply Hubbard formalism without strict
restriction on $t$ and $J$ values in $t-J$ model. According to Angle
Resolved Photoemission Electron Spectroscopy (ARPES) the bandwidth $W
\approx 3J$ for optimally doped HTSC \cite{DagottoRMP}. In undoped
insulating Sr$_2$CuO$_2$Cl$_2$, $W \sim J$, in agreement with Monte-Carlo
and Exact Diagonalization simulations \cite{DagottoRMP}, and the
valence-band dispersion has a maximum located near ($\pi/2$,$\pi/2$)
\cite{LaRosa}. The value of bare hopping between the first neighbors
$t_1=78$ meV is known from comparison with ARPES \cite{EreminBand}. We
stress the importance of $t_2$ and $t_3$ hopping integrals, that expand our
model and strongly affect on the Fermi surface form. The value of
superexchange interaction between the nearest Cu neighbors for
La$_{2-x}$Sr$_x$CuO$_4$ is known from neutron scattering data $J$=135 meV
\cite{HaydenLaSCO}. These arguments enforces us to study the opposite region
of values ($t_1, t_2, t_3 < J$), compared to usually used in celebrated
$t-J$ model.

The hoppings $t_{ij}$ are strongly affected by electron and especially by
spin-spin correlations, resulting to {\it effective} values of hoppings
\begin{equation}
t_{n}^{\mathit{eff}}=t_n P_{pd} \left[1+\frac{\langle {\bf S}_i {\bf S}_j
\rangle_n } {P_{pd}^2}\right]
\end{equation}
for $n$=$|j-i|$-th neighbor. $P_{pd}$=$(1+x)/2$ is Hubbard electronic
reduction factor. The spin-spin correlation functions $\langle {\bf S}_i
{\bf S}_j \rangle$ have been calculated in a variety of approaches
\cite{SpinCorrApproach} and numerical methods \cite{DagottoRMP} within the
2D $t-J$ model. The most notable result is that $\langle {\bf S}_i {\bf S}_j
\rangle_1 \approx -0.333$ in undoped antiferromagnet at zero temperature.
These calculations suggest also that with doping the spin-spin correlations
goes down first slowly in lightly doped and underdoped HTSC oxides and then
rapidly decreases for optimally doped and overdoped regime. Therefore, for
$x=1/8$ we set $\langle {\bf S}_i {\bf S}_j\rangle_1 = - P_{pd}^2 \approx -
0.3164$ so that $t_{1}^{\mathit{eff}}=0$. Strong antiferromagnetic
correlations block the hopping between the first neighbors so that the order
is {\it undamped} \cite{EreminRigamonti}. Our initial Fermi surface with
dispersion
\begin{eqnarray} \label{teff}
\varepsilon_{\bf k}^{\mathit{eff}} = 4t_{2}^{\mathit{eff}} \cos {k}_x \cos
{k}_y +2t_{3}^{\mathit{eff}} ( \cos 2{k}_x +\cos 2{k}_y),
\end{eqnarray}
has pockets centered around $(\pi/2,\pi/2)$ and is similar to that observed
by LaRosa {\it et al.} \cite{LaRosa} in Sr$_2$CuO$_2$Cl$_2$.

To select the instability wavevector for charge subsystem we have calculated
the charge susceptibility, using the expressions given in \cite{EreminSusc}.
The results are shown on Fig. 1. It is seen, that at low temperatures charge
susceptibility has a maximum near $(\pi/2,\pi)$, in remarkable agreement
with neutron scattering experiments \cite{Tranquada45}. With increasing
temperature the charge susceptibility shows a peak around ${\bf
Q}=(\pi,\pi)$. These results are stable with respect to moderate variation
of $G$ value.

To formulate the self-consistent system of equations we use Green's function
method and apply Roth's decoupling scheme \cite{Roth}. The elementary
excitations spectrum of the Hamiltonian (\ref{Hamiltonian}) is given by the
equation
\begin{equation}
{\mathrm det} \left| \begin{tabular}{llll}
\mbox{\hspace{1mm}}$ \varepsilon_{\bf k}^{\uparrow}-\mu$ & \mbox{\hspace{6mm}}$\eta_{\bf
k,Q}^{\uparrow}$ \nonumber \\
\mbox{\hspace{0mm}}$ \eta_{\bf k+Q,-Q}^{\uparrow}$ & \mbox{\hspace{2mm}}$\varepsilon _{\bf k+Q}^{\uparrow}-\mu$
\end{tabular}
\right| = 0. \label{matrix}
\end{equation}

The quasiparticle dispersion relation has the form
\begin{eqnarray} \label{epsilonk}
\varepsilon^{\sigma}_{\bf k}={P_{pd}} \sum_{i,j} t_{ij}
\left[1+\frac{\left<{\bf S}_i {\bf S}_j\right>_n }{P^2_{pd}}
\right] e^{i {\bf k R}_{ij} }
-\frac {1}{2NP_{pd}}\sum_{\bf k'} {\Big [}\left(J_{\bf k'-k}-2t_{\bf k'}
\right) \left< X^{2 \tilde \sigma}_{\bf k'} X^{\tilde \sigma 2}_{\bf k'}
\right> + 2G_{\bf k'-k}\left< X^{2\sigma}_{\bf k'}X^{\sigma 2}_{\bf
k'}\right> {\Big ]} ,
\end{eqnarray}
where $ t_{\bf k} = 2t_1 (\cos{k}_x +\cos{k}_y )+ 4t_{2} \cos {k}_x \cos
{k}_y + 2t_{3} ( \cos 2{k}_x +\cos 2{k}_y )$. The order parameter
$\eta_{\bf k,Q}^{\sigma}$ is responsible for abnormal properties of HTSC
layered copper oxides due to instabilities. It is caused by short ranged
interactions and given by

\begin{eqnarray} \label{etaDef}
\eta^{\sigma}_{\bf k,Q}&=&\left[t_{\bf k+Q}-\frac{\sum \left<{\bf S}_i
{\bf S}_j\right> t_{ij} e^{i{\bf k R}_{ij}}}{P_{pd}^2}+\frac{\left(
\sum\left<{\bf S}_i {\bf S}_j\right> J_{ij} e^{i {\bf k R}_{ij}}\right)
_{\bf k=0}}{2P_{pd}^2}-\frac{1}{2} J_{\bf Q} \right]\left< \frac{1}{2}
e_{\bf Q} + \sigma s^z_{\bf Q}\right> +
 \nonumber \\ 
&+&\frac {1}{2NP^2_{pd}}\sum_{\bf k'}{\Big [}\left( J_{\bf k'-k}-2t_{\bf k'}
\right) \left< X^{2 \tilde \sigma}_{\bf k'} X^{\tilde \sigma 2}_{\bf k'}
\right> + 2G_{\bf k'-k} \left< X^{2 \sigma}_{\bf k'} X^{\sigma 2}_{\bf k'}
\right> {\Big ]} \left< \frac {1}{2} e_{\bf Q} + \sigma s^z_{\bf Q} \right>-
 \nonumber \\ 
&-& \frac {1}{2NP_{pd}} \sum_{\bf k'}{\Big [}\left(J_{\bf k'-k}-2t_{\bf k'}
\right) \left< X^{2 \tilde \sigma}_{\bf k'+Q} X^{\tilde \sigma 2}_{\bf k'}
\right> + 2G_{\bf k'-k} \left< X^{2 \sigma}_{\bf k'+Q} X^{\sigma 2}_{\bf k'}
\right> {\Big ]} + G_{\bf Q} e_{\bf Q},
\end{eqnarray}
where $J_{\bf k} = 2 J \left(\cos k_x + \cos k_y \right)$, $G_{\bf k} = 2 G
\left(\cos k_x +\cos k_y \right)$ and $\sigma = \pm$ is spin index.
The latter contribution in Eq. (\ref{etaDef}), $G_Q e_Q$, will be important
and we note, that $G_Q/G=-4$ is negative due to instability vector ${\bf
Q}=(\pi,\pi)$. The eigenvalues of the matrix (\ref{matrix}) are given by
\begin{equation} \label{E12}
E_{1{\bf k},2\bf k}^{\sigma}=\frac{1}{2}
\left( \varepsilon^{\sigma}_{\bf k+Q} + \varepsilon^{\sigma}_{\bf k} \right)
 \frac {}{} - \mu \pm \frac{1}{2} E^{\sigma}_{12}
\end{equation}
where
\begin{equation} \label{E12sqrt}
E^{\sigma}_{12}=\sqrt{\left( \varepsilon^{\sigma}_{\bf k+Q}
- \varepsilon^{\sigma}_{\bf k} \right)^2
+ 4 \eta^{\sigma}_{\bf k+Q,Q} \eta^{\sigma}_{\bf k,Q} }.
\end{equation}
It is seen that in the $t_{1}^{\mathit{eff}}=0$ case $\eta^{\sigma}_{\bf
k,Q} = \eta^{\sigma}_{\bf k+Q,-Q}$ and the quasiparticle energy is real,
i.e. the regime realized has no damping.
\begin{equation} \label{eQdef}
e_{\bf Q} = \frac {1}{N} \sum \delta_i e^{i {\bf Q R}_i},
\mbox{\hspace{5mm}}
s_{\bf Q}^z = \frac {1}{N} \sum s_i^z e^{ i {\bf Q R}_i}
\end{equation}
are Fourier amplitudes of CDW and Spin Density Wave (SDW) respectively and
can be expressed via thermodynamic averages as
\begin{equation}  \label{eQGreen}
\left< e_{\bf Q} \right> = \frac {1}{2N} \sum_{\bf k}
\left< X^{2 \uparrow}_{\bf k+Q} X^{\uparrow 2}_{\bf k} +
 X^{2 \downarrow}_{\bf k+Q} X^{\downarrow 2}_{\bf k} \right>,
\end{equation}
\begin{equation}  \label{sQGreen}
\left< s_{\bf Q}^z \right> = \frac {1}{2N} \sum_{\bf k}
\left< X^{\uparrow 2}_{\bf k} X^{2 \uparrow}_{\bf k+Q} -
 X^{\downarrow 2}_{\bf k} X^{2 \downarrow}_{\bf k+Q} \right>.
\end{equation}
We note, that the $id$-wave order does not contribute to expectation values
of CDW $(e_{\bf Q})$ and SDW $(s_{\bf Q}^z)$.

The relevant thermal average have the following form:
\begin{equation}  \label{XkQXk}
\left< X^{2 \sigma}_{\bf k+Q} X^{\sigma 2}_{\bf k}\right> =
\frac {P_{pd}\mbox{\hspace{0.3mm}}\eta^{\sigma}_{\bf k,Q}}
 {E^{\sigma}_{12}}
\left[ f(E_{1\bf k}^{\sigma})-\frac {}{} f(E_{2\bf k}^{\sigma})\right].
\end{equation}

In the $t_1^{\mathit{eff}}=0$, $t_2^{\mathit{eff}}, t_3^{\mathit{eff}}\ll
G$ limit suggestions on solution of system of equations
(\ref{etaDef}-\ref{XkQXk}) may be obtained even analytically. Suppose, that
we have non-negligible CDW amplitude $e_Q$. Then, due to large $G$,
compared to hoppings, value, the contribution $G_Q e_Q $ will dominate in
(\ref{etaDef}). Next, we note, that the $\eta$ order parameter will
dominate in Eq. (\ref{E12sqrt}) and hence $E_{12}^{\sigma} \approx
|2\eta_{\bf k,Q}^{\sigma}|$. Therefore the absolute value of thermal
average (\ref{XkQXk}) is equal to $\frac {1}{2} P_{pd}$. Substituting this
result to Eqs. (\ref{eQGreen}, \ref{sQGreen}) we obtain, that the SDW
amplitude is $s_Q = 0$ and CDW amplitude $e_Q=\frac {1}{2} P_{pd}$.

Despite many attempts to calculate the numerical values of hoppings
$t_2^{\mathit{eff}}$ and $t_3^{\mathit{eff}},$ their values are precisely
still unknown together with the spin-spin correlation functions between the
next-next nearest neighbors. The only known result from ARPES experiments
and exact numerical simulations within 2D $t-J$ model is that the bandwidth
$W \sim J$ for undoped antiferromagnet, that gives a restriction to the
$t_2^{\mathit{eff}}$ and $t_3^{\mathit{eff}}$ values.  We set the values of
{\it effective} hoppings $|t_2^{\mathit{eff}}|=10$ meV and
$t_3^{\mathit{eff}}= 20$ meV, that are, in our opinion, the most relevant
values for La$_{2-x}$Sr$_x$CuO$_4$ with $x=1/8$. For the spin-spin
correlation function between the second neighbors we set $\langle {\bf S}_i
{\bf S}_j\rangle_2 = 0.15$, due to reasons described above. The value and
even the sign of next-next nearest neighbors spin-spin correlation function
is completely unknown, therefore we set $\langle {\bf S}_i {\bf S}_j
\rangle_3 = 0$. $F_{i}\approx$ 4~meV was estimated in \cite{EreminSigmund}.
The value of {\it screened} Coulomb repulsion at $x \approx 1/8$ is
$G\approx 75$ meV \cite{EreminSusc}. We have solved the system of equations
(\ref{etaDef}-\ref{XkQXk}) with this set of parameters. We have obtained,
that in the low temperature limit the CDW amplitude $e_Q=x$, where $x$ is
the number of holes in CuO$_2$ plane due to Sr doping. In this case the {\it
undamped} CDW completely destroy superconductivity at $x=1/8$ in
La$_{2-x}$Sr$_x$CuO$_4$ and are responsible for drastic change in
temperature dependencies of various physical quantities. The calculated
wavevector dependence of $\eta_{\bf k}^{\sigma}$ order parameter at T$=50$~K
is presented in Fig. \ref{Figure2}. It is seen, that its value is mainly
determined by wavevector independent CDW contribution $G_Q e_Q$. The
temperature dependence of CDW amplitude $e_Q$ determined by temperature
dependence of thermal average (\ref{XkQXk}), which are, in turn, determined
by mean field approximation via Fermi distribution functions $f(E_{1{\bf
k}}^{\sigma}), f(E_{2{\bf k}}^{\sigma})$. The calculated CDW closure
temperature is T$_{CDW}=350$~K.

The described above scenario explains the reason of anomaly at $x=1/8$. At
this doping level conventional CDW are allowed (without damping) and they
suppress superconductivity. According to our calculation CDW regime appeared
at $x=1/8$ far above T$_{\mathrm C}$. The experimental observation of
critical temperature T$_{d2} \simeq$ 138~K in
La$_{2-x-y}$Eu$_y$Sr$_x$CuO$_4$ with $x=1/8$ at which the copper relaxation
rate $1/T_1$ suddenly goes down \cite{MatsumuraJpn} can be naturally
interpreted as a transition related to CDW state. With lowering temperature
the pinning process due to freezing of the spin fluctuation becomes
important and charge-spin ordering tends to be quasi-static. In Fig.
\ref{Figure3} we show our variant of copper - oxygen singles ordering in
plane which we expect to be appeared in La$_{2-x-y}$Re$_y$Sr$_x$CuO$_4$
(Re=Nd and Eu) below T $\simeq$ 9~K. The displayed stripe pattern has many
common features with that suggested by Tranquada {\it et al.}
\cite{Tranquada45}. The distinction is that our picture corresponds to the
so-called in-phase domain walls in contrast to antiphase domain proposed in
\cite{Tranquada45}. As one can see our model explains the observed magnetic
superlattice peaks of the type (1/2,$\pm \epsilon$), (1/2,0) (in $2\pi$
units) and charge-ordered peaks at (2$\epsilon$,0,0) \cite{Tranquada45},
too. In addition it is naturally follows from widely accepted copper-oxygen
singlet formation picture \cite{ZhangRice}, which provides the minimum in
the energy of the exchange interactions between copper - oxygen and
copper-copper spins \cite{EreminSigmund}. The last term in
(\ref{Hamiltonian}) plays crucial role in this aspect. It was pointed in
Ref. \onlinecite{MatsumuraJpn} that in-phase domain even better reproduce
the observed high frequency tail in NQR spectra in La$_{2-x}$Ba$_x$CuO$_4$
with $x=1/8$ at low temperatures compared to anti-phase domain model
\cite{Tranquada45}. This conclusion is consistent with the ferromagnetic
stripe ordering suggestion \cite{Lavrov} based on strong sensitivity to
magnetic fields. The Coulomb term ($G$) is important, since without it there
is no CDW-like instability. This statement is in agreement with conclusion
of \cite{HellbergManousakis} that pure $t-J$ model does not give any
stripes.

Finally we comment the results of huge isotope shift changes with Sr doping
in La$_{2-x}$Sr$_x$CuO$_4$ \cite{Zhao}. According to our findings the
competition between $d$-wave superconductivity and CDW is stronger, compared
to competition between $d$-wave SC and order, caused by short range
interactions. Moreover, the latter type of competition becomes stronger with
decreasing of doping. Thus we are able to explain the gradual increase of
isotope effect with decreasing of Sr doping in underdoped regime together
with nearly divergent behavior of isotope exponent near $x=1/8$ anomaly in
La$_{2-x}$Sr$_x$CuO$_4$.

In summary, the calculated charge susceptibility shows a maximum near
$(\pi,\pi)$ at intermediate temperatures and near $(\pi,\pi/2)$ as
temperature approaches zero, in agreement with neutron scattering
experiments. We have shown, that in two-dimensional systems with narrow
bands, so that the hoppings $t_2^{\mathit{eff}}, t_3^{\mathit{eff}} < G$,
$t_{1}^{\mathit{eff}} = 0 $ due to strong antiferromagnetic correlations,
and with instability antiferromagnetic wave vector $Q=(\pi,\pi)$, due to
large Coulomb repulsion between the first neighbors $G$ the CDW are formed.
For physically realistic values $t_2^{\mathit{eff}}, t_3^{\mathit{eff}}$ at
$x=1/8$ in La$_{2-x}$Sr$_x$CuO$_4$ the CDW amplitude $e_Q$ = $x$, where $x$
is Sr content. Based on the calculated charge susceptibility peaks position
the in-phase domain structure as a candidate for "stripe" picture is
proposed. In our opinion, this scenario sheds light on drastic change of
various physical quantities at "magic" value of Sr content $x=1/8$ in
La$_{2-x}$Sr$_x$CuO$_4$.

This work was supported in part by Russian State Science and Technology
Program "Superconductivity" Grant No. 98014-2 and CRDF Rec 007. I.A.L.
thanks INTAS YSF 2001/2-45 for support.

\begin{figure}  [tbp]
  \centering
  \includegraphics[height=0.5\linewidth,width=0.8\linewidth]{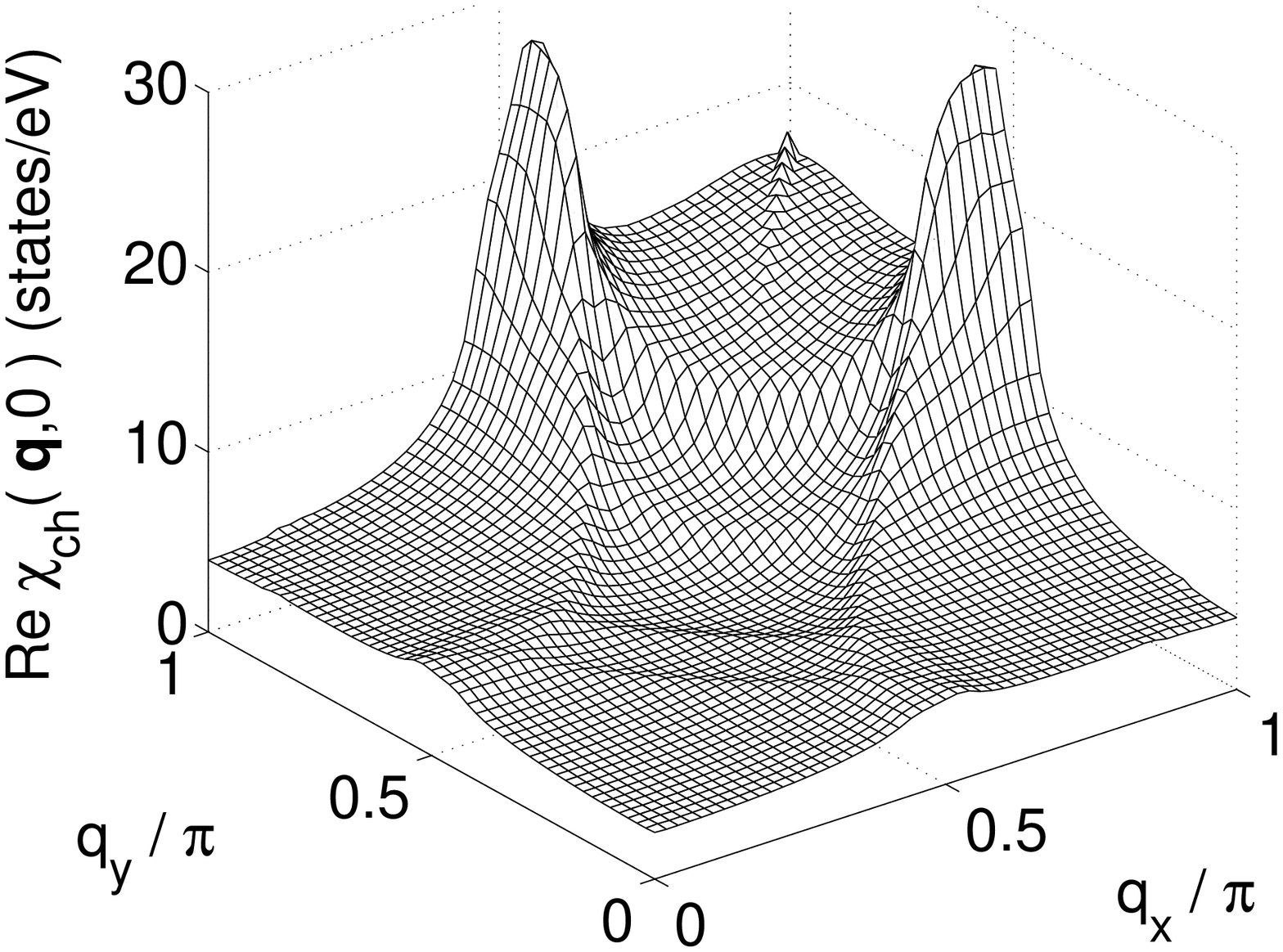}
\caption{The calculated real part of charge susceptibility at T = 10 K with
the parameters as described in the text. }
\label{Figure1}
\end{figure}

\begin{figure}  [tbp]
  \centering \includegraphics[height=0.5\linewidth,width=0.8\linewidth]
{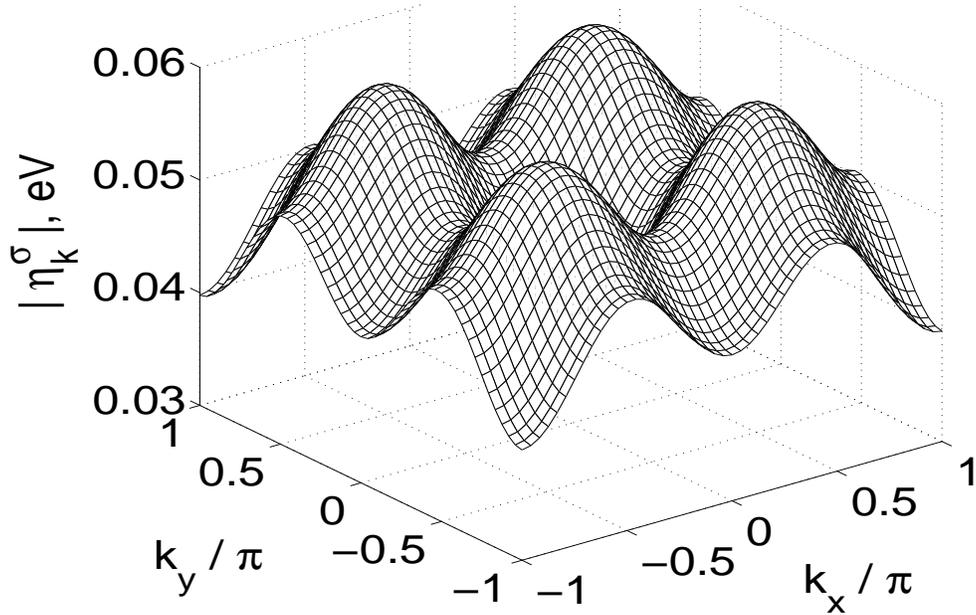} \caption{The calculated wavevector dependence of
$\eta_k^{\sigma}$ order parameter. The main value of $\eta_k^{\sigma}$
originates by wavevector independent Coulomb $\times$ CDW amplitude
contribution $G_Q e_Q$.}
\label{Figure2}
\end{figure}

\begin{figure}  [tbp]
  \centering \includegraphics[height=0.3\linewidth,width=\linewidth]
  {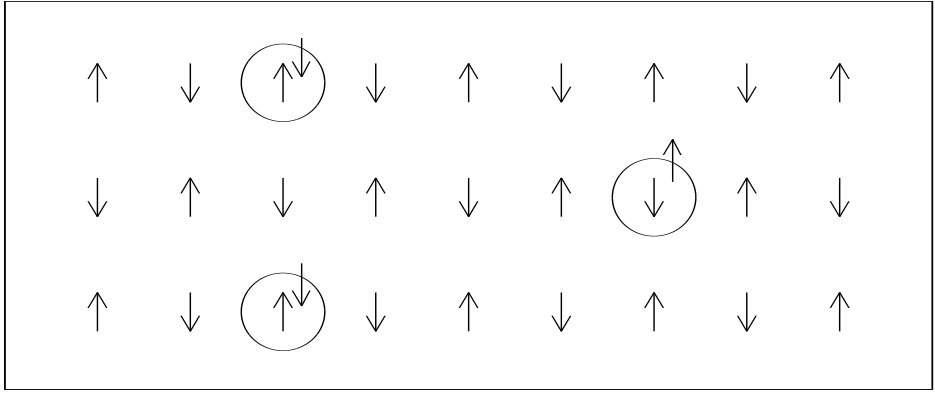}
\caption{Sketch of in-phase "stripe" picture.
The circles correspond to oxygen holes.}
\label{Figure3}
\end{figure}
\end{document}